\documentclass{aa}
\usepackage{graphicx}
\usepackage[normalem]{ulem}
\usepackage{txfonts}
\begin{document}

   \title{Stellar Magnetic Activity and ``Butterfly Diagram" of Kepler-63}

   \author{Y. Netto,
          \inst{1}
          \and
          A. Valio\inst{1}
    % ORCID: 0000-0002-1671-8370
          }

   \institute{ CRAAM, Mackenzie Presbyterian University,
              Rua da Consolacao, 896, Sao Paulo, Brazil\\
              \email{dirceuyuri@hotmail.com} }

   \date{Received ... , ; accepted  ...  }

% \abstract{}{}{}{}{} 
% 5 {} token are mandatory
 
  \abstract
   {The study of young solar type stars is fundamental for a better understanding of the magnetic activity of the Sun. Most commonly, this activity manifests itself in the form of spots and faculae. As a planet in transit crosses in front of its host star, a dark spot on the stellar surface may be occulted, causing a detectable variation in the light curve. Kepler-63 is a young solar-like star with an age of only 210 Myr that exhibit photometric variations compatible with spot signatures.  Since its orbiting planet is in an almost polar orbit, different latitudes of the star can be probed by the method of spot transit mapping.}
   {The goal of this work is to characterise the spots of Kepler-63 and thus decipher the behaviour of the young Sun. Because of the high obliquity of the planetary orbit, the spots latitudinal distribution and thus its differential rotation may be determined. }
   {A total of 150 transits of Kepler-63b were observed in the short cadence light curve, corresponding to a total duration of about 4 years. Each transit light curve was fit by a model that simulates planetary transits and allows the inclusion of starspots on the surface of the host star. This enables the physical characterisation of the spots size, intensity, and location. We determine the spot position in a reference frame that rotates with the star, and thus obtain the latitudinal distribution of the spots.}
   {A total of 297 spots were fit and their sizes, intensities, and positions determined. The spots longitude and latitude were calculated on a reference frame that rotates with the star. The latitude distribution of spots exhibits a bimodality with a lack of spots around $34^\circ$. Moreover, the size of spots tend to be larger close to the equator, decreasing toward the latitude distribution gap, and then increasing again toward the poles. The high latitude spots dominate the magnetic cycle of Kepler-63. For a mean stellar rotation period of 5.400d, 59 spots were found at approximately the same longitude and latitude on a later transit. Some of these spots were detected 8 transits later, showing the existence of spots with lifetimes of at least 75d. }
   {Due to the geometry of the Kepler-63 system, we were able to build a starspot ``butterfly diagram", similar to that of sunspots. It was also possible to infer Kepler-63 differential rotation from the presence of spots at different latitudes. This star was found to rotate almost rigidly with a period of 5.400d and relative shear close to 0.01\% for latitudes less than $34^\circ$, whereas the high latitudes do not follow a well behaved pattern.}

   \maketitle
%________________________________________________________________

\section{Introduction}
%_________________________________________________________________
Sunspots are regions of high concentration of magnetic fields, being cooler, and therefore darker than the surrounding photosphere. Just like the Sun, dark spots appear on the surface of stars \citep{Strassmeier09}. 
When an orbiting planet eclipses a star, there is a chance that it will cross in front of a solar-like spot on the stellar surface, causing a detectable variation in the transit light curve. Based on this property, \cite{Silva03} developed a method that allows to detect spots as small as 0.2 planetary radii. Furthermore, using this method it is possible to infer properties of individual starspots on the occulted transit band, such as size, intensity, and position. This technique has already been applied to HD 209459 \citep{Silva03}, CoRoT-2 \citep{SilvaValio10, SilvaValioLanza11}, Kepler-17 \citep{Valio17}, and Kepler-71 \citep{Zaleski19}. Using the transit technique to investigate the characteristics of starspots, and considering the rotation of the star it is possible to measure the stellar differential rotation \citep{Valio13}. Several transit-starspots models have been developed by the exoplanet community over the years, for example: PRISM: \citet{Tregloan13, Tregloan15, Tregloan18}; SOAP-T: \citet{Oshagh13}; spotrod: \citet{Beky14}; KSint: \citet{Montalto14}; ellc: \citet{Maxted16}; StarSim: \citet{Herrero16}; PyTranSpot: \citet{Juvan18}.

In the Sun, differential rotation is an important mechanism for the solar dynamo \citep{Parker55, Schrijver00}. Other key ingredients related to dynamo models that can be inferred through the study of starspots are: butterfly diagram \citep{SanchisOjeda11, Morris17, Hackman19}, period of magnetic cycle \citep{EstrelaValio16}, and starspot lifetime \citep{Berdyugina05}. Thus by studying a sample of active stars we hope to be able to better understand the stellar dynamo.

Kepler-63 is a young, solar-like star that exhibits high activity. It has an estimated age of (210 $\pm$ 45) Myr, a mean rotation period of $\sim$ 5.4d, and is orbited by a giant planet with a polar orbit of 9.43d period. Kepler-63b polar orbit \citep{Ojeda13} allows to map the spots  at different latitudes occulted during the transits. \cite{Ojeda13} characterised this system by analysing $\sim$ 900 days of Kepler data available at that time. During that time, 96 planetary transit were observed. More recently the magnetic activity cycle of Kepler-63 was characterised by \cite{EstrelaValio16}. The authors found a magnetic cycle period of 1.27yr. Here we extend the analysis of \citet{Ojeda13} by analysing the complete Kepler data of high-precision photometric time series covering $\sim$ 1400 days of Kepler-63 system using for the first time the transit method described by \cite{Silva03} for a star with a planet in a polar orbit. This type of orbit provides information on the spot latitudes.

The next section gives an overview about the observation of the Kepler-63 system. Section 3 describes the spot model used in this work, as well as the physical parameters obtained for the modelled spots. Section 4 analyses the starspots latitude temporal evolution, whereas Section 5 describes the differential rotation. Finally, Section 6 presents the conclusions.

\section{Observations}

The Kepler telescope was developed to detect planetary transits of stars at a fixed field in the Cygnus constellation. The 95-cm telescope, that monitored approximately 150,000 stars, had to be rolled by 90$^\circ$ about its line of sight every three months to optimise solar panel efficiency. Therefore, the Kepler operation used to be divided in four quarters each year, separated by its rolls. Each target observed used to fall in different CCDs during different quarters. Kepler observations were performed with two cadences: long cadence (1765.5 s or 29.4 min) and short cadence (58.5 s or 0.975 min). Since we are interested in the detection of spots during planetary transits,  we analyse the photometry acquired in short cadence mode. The long cadence is used only to analyse the rotational modulation for stellar period determination which have timescale of the order of several days \citep{Lanza19}.

Our analysis was performed on the short cadence Presearch Data Conditioning (hereafter PDC) light curve time series of the latest data release (Data Release 25) of Kepler-63. The PDC tends to remove instrumental and systematic effects from the Simple Aperture Photometry (SAP) light curve while leaving the activity signatures unaffected \citep{Stumpe12, Stumpe14}.

\subsection{The Kepler-63 system}

\begin{figure}[t]
\centering
\includegraphics[width=8.8cm]{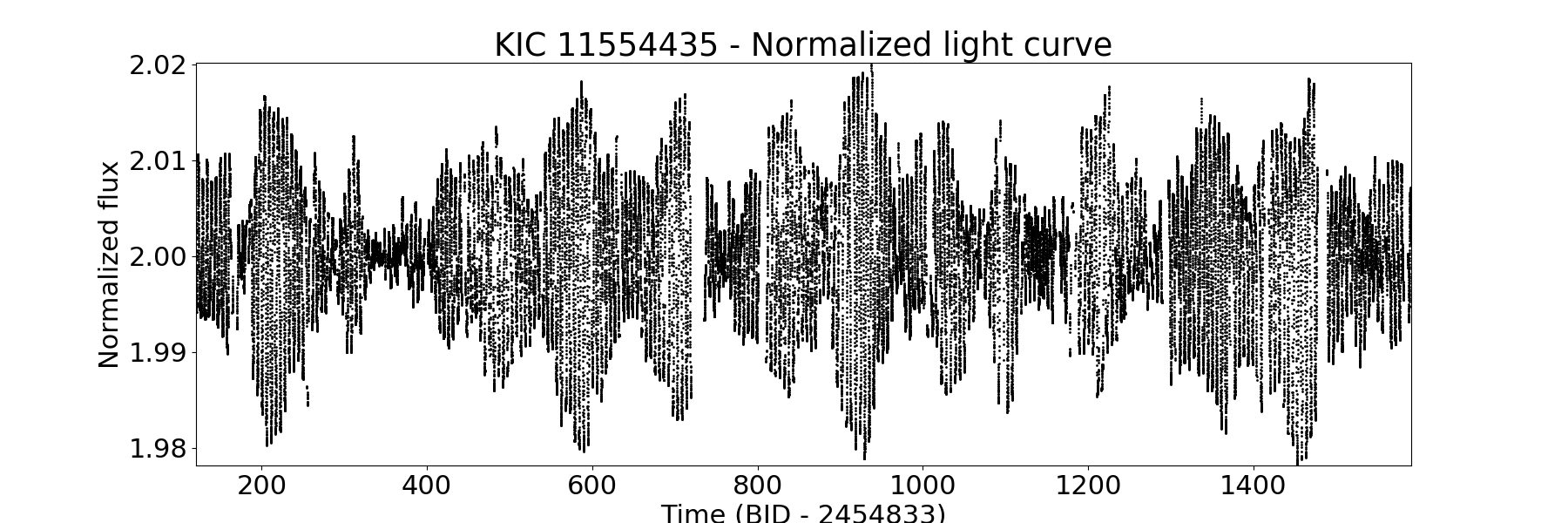} 
     \caption{Normalized Kepler-63 light curve covering $\sim$ 1400 days. The peak-to-peak photometric modulation of 4\% is due to the presence of spots on the stellar surface that rotate in and out of view (BJD - 2454833).}
\label{k63LC}
\end{figure}

Kepler-63 is a young solar type star, with approximately 210 Myr, that exhibits high activity. Figure~\ref{k63LC} shows the full 4 years Kepler-63 light curve marked by rotational modulations, with a peak-to-peak modulation of 4\%, clearly showing its magnetic activity. This star hosts a giant planet in a polar orbit with a period of 9.43 days. Details of the star, the planet, and its orbital parameters are listed on Table~\ref{param}. During the nearly 4 years of observation, a total of 150 transits were observed by the Kepler satellite.

\begin{figure}[t]
\centering
\includegraphics[width=9cm]{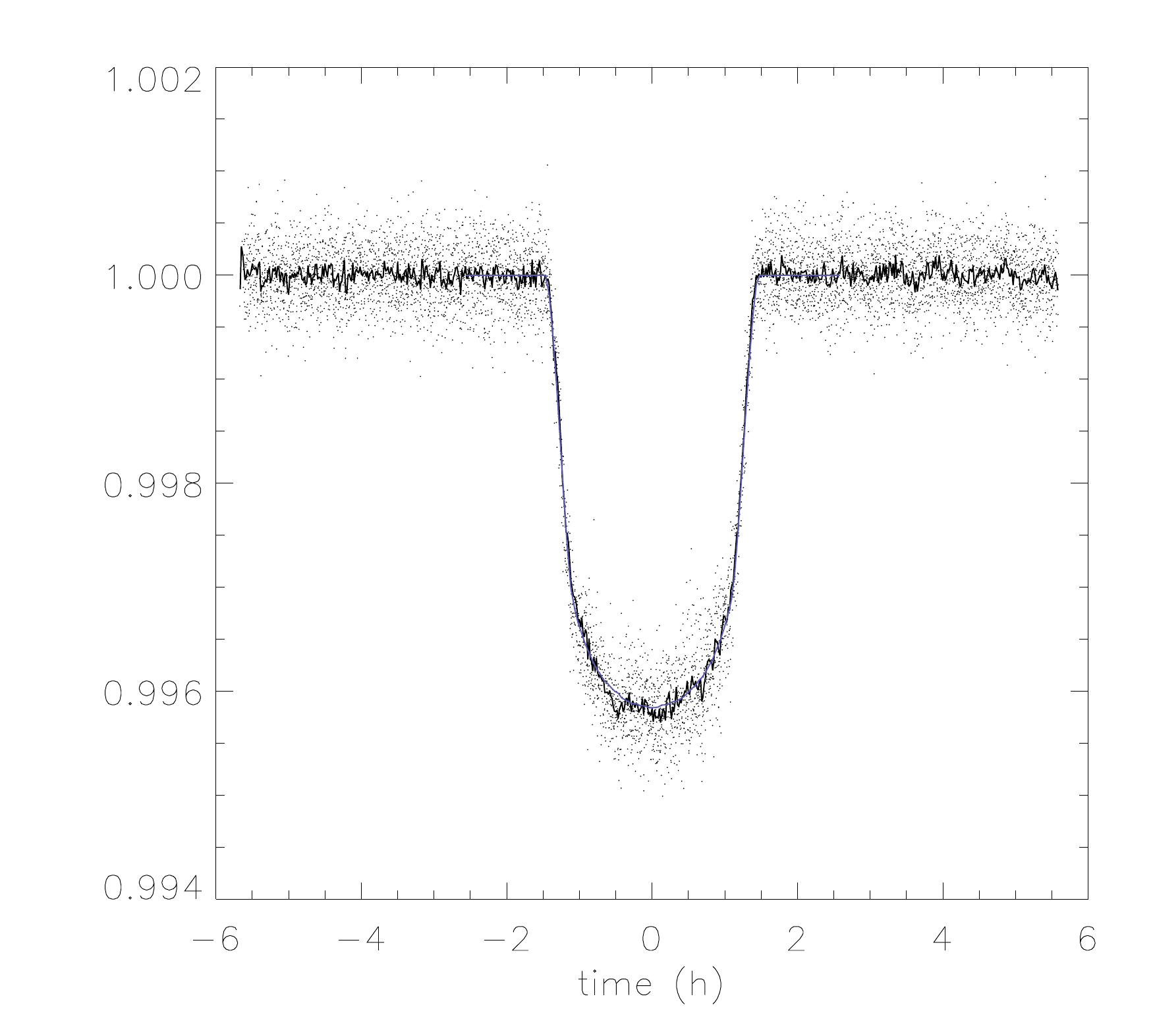} 
     \caption{10 deepest transits phased. Fitted spotless light curve in blue.}
\label{k63spotless}
\end{figure}

The spots influence on the transit light curve is twofold. First, the spots artificially make the transit shallower, thus underestimating the planet radius. Moreover, the spots near the limb affect the transit duration, altering the semi major axis determination. To better estimate the transit parameters, we considered the 10 deepest transits without any visible spot signature (Figure~\ref{k63spotless}). We then median-combined these transits (solid black curve in Figure~\ref{k63spotless}) and fit a spotless star model to them (red curve in Figure~\ref{k63spotless}), that yields the values for the radius, semi-major axis, and orbital inclination angle given in Table~\ref{param}. Also fitted by this model were the limb-darkening coefficient $\omega_1$ and $\omega_2$ using the expression presented in \cite{Brown01}:

\begin{equation}
\frac{I(\mu)}{I(1)} = 1 - \omega_1(1 - \mu) - \omega_2 (1 - \mu)^2
\label{obslimbo1}
\end{equation} 

\noindent where $\mu$ is the cosine of the angle between the line of sight and the normal to the local stellar surface. As can be seen, some values were slightly increased, the planet radius by \textbf{4\%} whereas the semi major axis by 1.2\% with respect to those used by \cite{Ojeda13}. 

\begin{table}[h]
\caption{Parameters of the Kepler-63 system, as presented in \cite{Ojeda13}, except those marked with $\ast$ that were modified for this work (see text).}
\tiny
\begin{center}
\begin{tabular}{lcl}
\hline\hline
Parameter &	Value & Ref. \\
\hline
\textit{Star} \\
\hline
Effective Temperature $T_{eff}$, (K),  & 5576 ($\pm$50) &	S13\\
Mass $M\star$, ($M\odot$) & 0.984 (-0.04, +0.035) &	S13\\
Radius $R\star$, ($R\odot$) &	0.901 (-0.022, +0.027) & S13\\
Rotation Period$^\ast$ $P_{star}$, ($days$) & 5.400 ($\pm$0.009) & NV19\\
Age, ($Myrs$) &	210 ($\pm$45) & S13\\
Limb-darkening coefficient $\mu_1$ & 0.31 ($\pm$0.04) & S13\\
Limb-darkening coefficient $\mu_2$ & 0.354 (-0.05, +0.07) & S13\\
Sky-projected Stellar Obliquity, ($deg$) &	-110 (-14, +22) & S13\\
Inclination of rotation axis, ($deg$) &	138 ($\pm$7) & S13\\
\\
\textit{Planet} \\
\hline
Mass $M_p$, ($M_{Jup}$) & 0.4 & S13\\
Radius$^\ast$ $R_p$, ($R_p/R\star$) & 0.0644 & NV19\\
Orbital Period, ($days$) & 9.4341505 ($\pm$ 1$\times$10$^{-6}$) & S13\\
Semi major axis$^\ast$, ($a/R\star$) & 19.35 & NV19\\
Orbital inclination angle $i$, ($deg$) & 87.806 (-0.019, +0.018) & S13\\
\hline
\end{tabular}
\tablebib{S13: ~\citet{Ojeda13}; NV19: present study.}
\label{param}
\end{center}
\end{table}

\section{The Spot Model}

\begin{figure}[h]
\centering
\includegraphics[width=5cm]{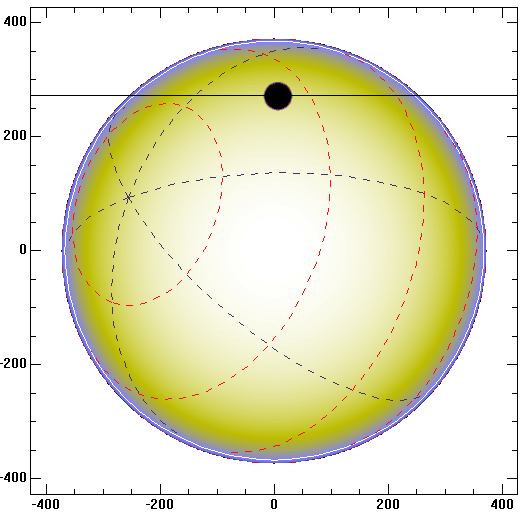} 
     \caption{Transit simulation of the planet crossing the Southern hemisphere of the star Kepler-63 in its nearly polar orbit. The stellar longitude (divided in 60$^\circ$) and latitude (divided in 30$^\circ$) are represented by blue lines and red lines, respectively. The black line represents the path of the planet (opaque disk).}.
\label{k63}
\end{figure}

We used the model described in \cite{Silva03} to simulate a synthesised star as a 2D image with limb darkening and the transiting planet as a dark disc with radius $R_p/R_{star}$, where $R_p$ is the radius of the planet and $R_{star}$ is the radius of the host star (see Figure~\ref{k63}). With the semi-major axis and inclination angle values, the orbit of the planet is calculated (black horizontal line in Figure~\ref{k63}) and assumed to occur in the Northern hemisphere of the star. The model assumes that the orbit is circular, that is, zero eccentricity.

The model also allows the inclusion of features on the stellar surface. During the transit of a planet in front of its star, it may occult solar-like spots or faculae on the stellar surface. This occultation produces small variations in the light curve. The model assumes that the spots are circular, and fits these small variations to obtain estimates of the physical parameters of starspots such as intensity, as a function of stellar intensity at disk center, $I_c$ (maximum value); size, or radius, as a function of planet radius, $R_p$; and position: latitude and longitude. When the spot is near the limb, the effect of foreshortening is also featured in the model.

\begin{figure}[t]
\centering
\includegraphics[width=8cm]{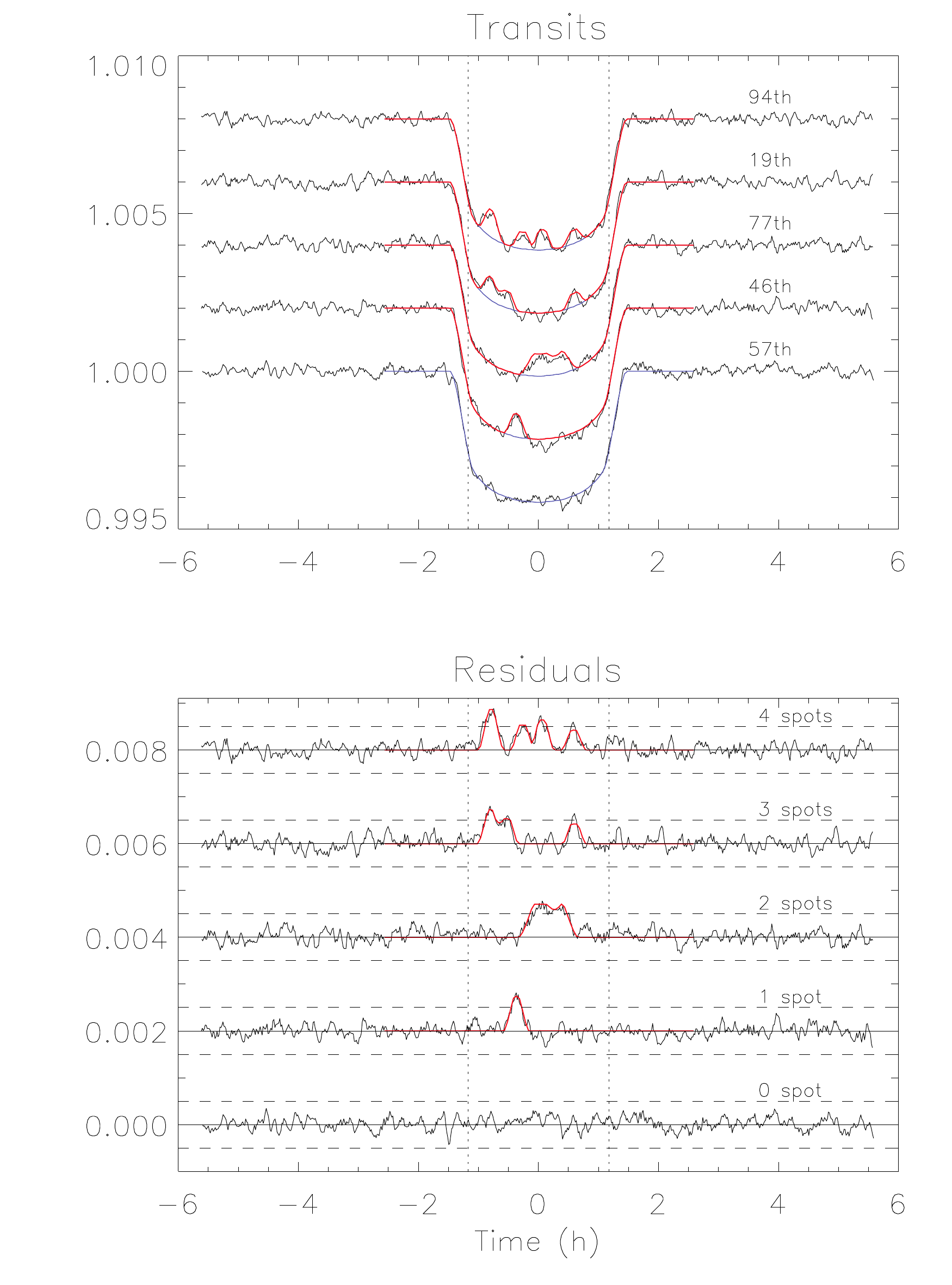} 
     \caption{Example of a fit to a transit. Top panel: Light curve of five transits, taking as the first transit as reference epoch (BJD = 2455010.79717), with the overploted blue curve that represents a transit in front of a spotless star, and the model with zero up to 4 spots (red curve). Bottom panel: Residuals after subtraction of the spotless model, with the best spot fit (red curve) from zero to four spots. Dashed horizontal lines represent the threshold for spot modelling that equals 10 times CDPP, and the dotted vertical lines at $\pm$ 1.18 h limits the fitting portion of the transit corresponding to  $\pm 70^\circ$ stellar longitudes.}
\label{transitcurv}
\end{figure}

This transit model described in \cite{Silva03} was applied to the 150 transits of Kepler-63b. Examples of five of these transits are shown in Figure~\ref{transitcurv}, smoothed every 5 data points. To better identity the small variations due to spots, the model light curve of a spotless star (blue curve in Figure~\ref{k63spotless} and top panel of Figure~\ref{transitcurv}) was subtracted from the observed transit light curve (black curve). The residuals resulting from this subtraction are plotted in the bottom panel of Figure~\ref{transitcurv}, for the five transits with zero, 1, 2, 3, or 4  modelled spots (red curve). A maximum of 4 spots per transit were fit and only the variations larger than 500ppm, that is 10 times the Combined Differential Photometric Precision (CDPP) \citep{Christiansen12}, of the data were considered (horizontal dashed lines in the bottom panel of Figure~\ref{transitcurv}). Note that no data outside of the transit rise above or below 10 times the CDPP value. 
Only spots within longitudes of $\pm 70^\circ$, which correspond to  $\pm$ 1.18 h in this case, are considered because of the difficulties in fitting spots at planet ingress and egress, where the transit light curve is very steep \citep{SilvaValio10}. This procedure resulted in a total detection of 297 spots.

\subsection{Spot longitude and latitude}

\begin{figure}[h]
\centering
\includegraphics[width=8cm]{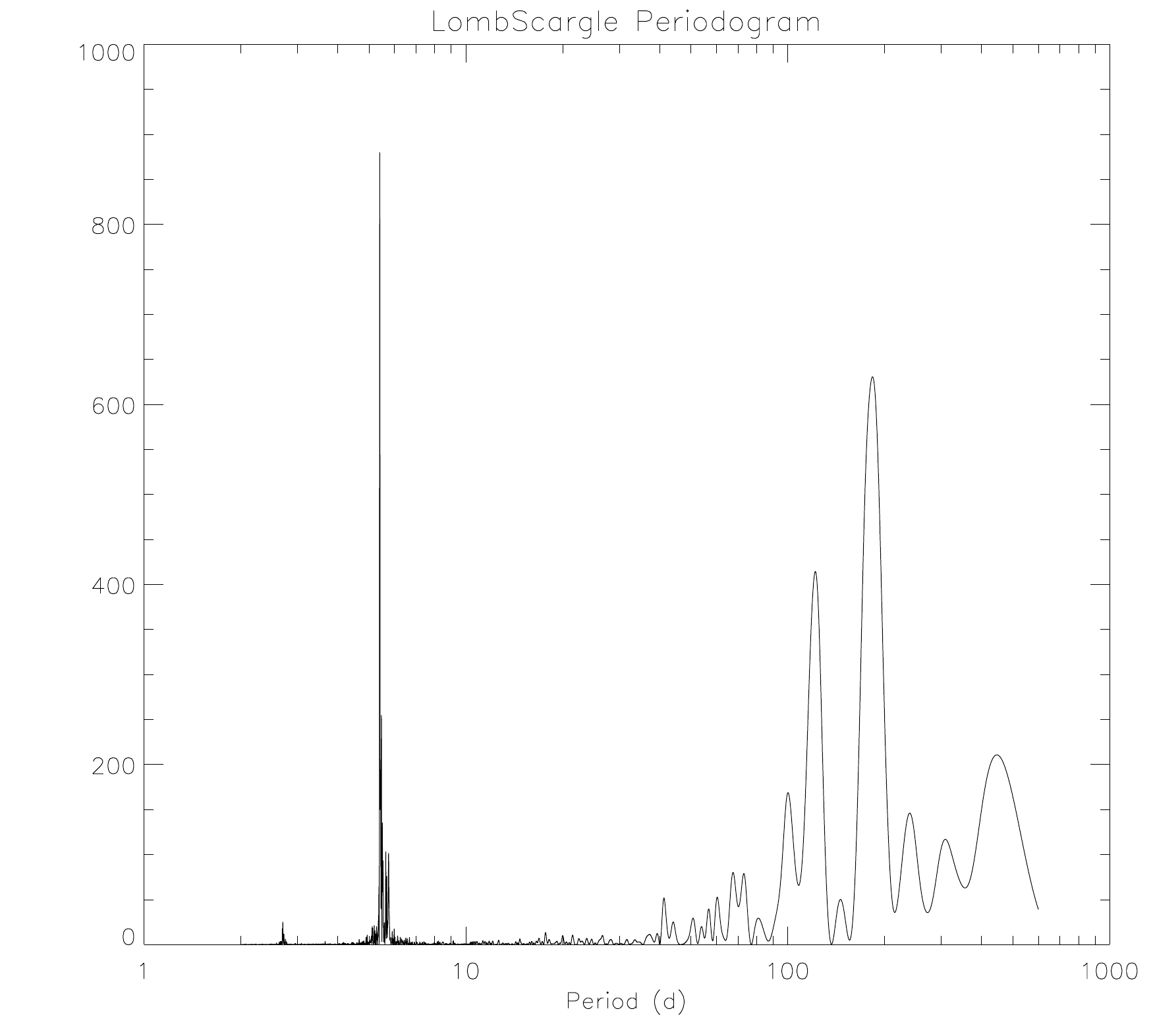} 
     \caption{Lomb-Scargle periodogram of the light curve of Kepler-63. The average rotation of the star is the main peak of the power spectrum.}
\label{fig:LS}
\end{figure}

Since, in the case of the Kepler-63 system, the planet orbit is not aligned with the stellar equator (Figure~\ref{k63}), the position of the spots obtained from the model has to be rotated to a position in an inclined reference frame that is aligned with the stellar rotation axis. For that, we need the mean rotation period of the star. As a first approximation, we use the one obtained from the maximum of the Lomb-Scargle periodogram \citep{Lomb76,Scargle82}, that corresponds to a rotation period of ($5.400 \pm 0.009$  days, with the standard deviation obtained by a Gaussian fit to the peak of the power spectrum (Figure~\ref{fig:LS}), very close to the 5.401 days found by \cite{Ojeda13}.

To determine the spots actual longitude and latitude on the surface of the star, the coordinates need to be given with respect to a reference system that rotates with the star. Since the spot model considers that the planet orbit is aligned with the stellar equator, the planetary transit shadows a ``fixed" stellar latitude as seen from Earth. For Kepler-63b, according to the geometry of the orbit, $lat_{topo}=\arcsin (a \cdot\cos(i)) = 48^\circ$, where $a$ is the semi major axis and $i$ the inclination angle of the orbit, respectively.

We then used a 3D rotation matrix taking into consideration the inclination of the rotation axis, sky-projected stellar obliquity and rotation, given in Table~\ref{param} to determine the spots position that rotates with the star. It also consider the rotation of the star varying in time given by $\Omega t = 2\pi/P_{rot}\cdot k \cdot P_{orb}$, where $k$ represents the {\it k-th} transit, $P_{rot}$ is the rotation period and $P_{orb}$ is the orbital period. First we rotate about the $x$ axis by the $\psi$ angle, then we rotate about the $y'$ axis by the $\theta$ angle, and finally we rotate about the $z''$ axis with an angle of $\Omega t$, where $\psi$ is the sky-projected stellar obliquity and $\theta$ is the inclination of rotation axis (given in Table~\ref{param}).

\begin{figure}[h]
\centering
\includegraphics[width=9cm]{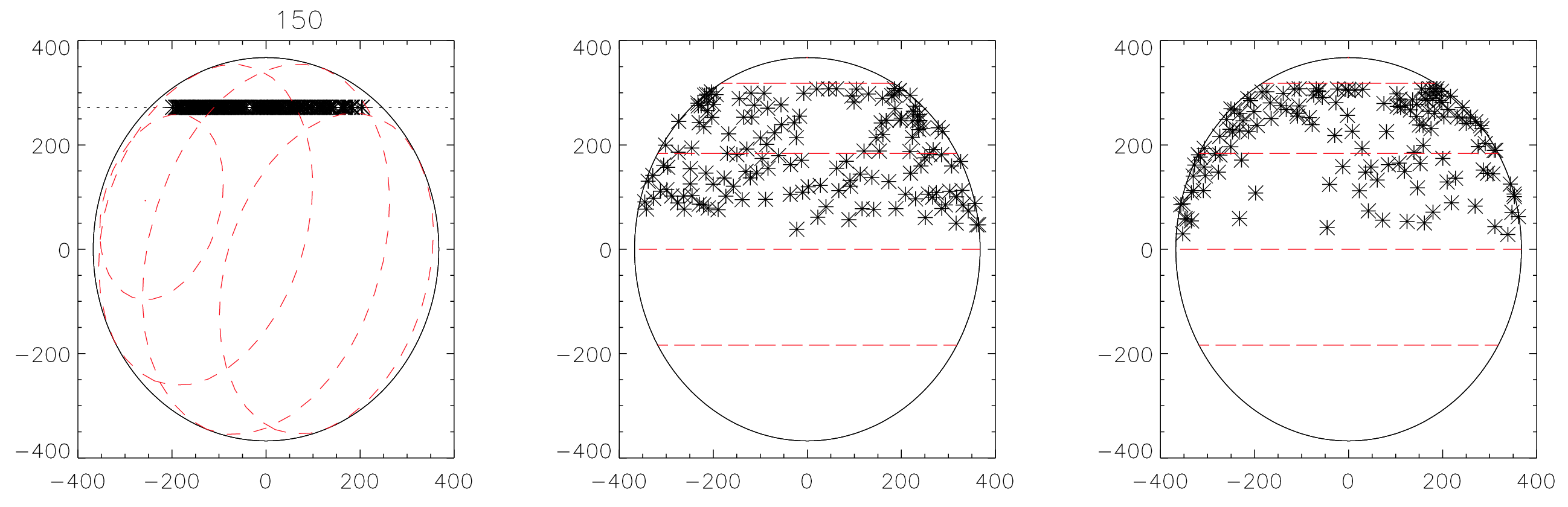} 
     \caption{Synthesised image of Kepler-63 with temporal overlap of all spots. Left: Referential frame seen from Earth. Middle: The spots with their location (positive longitudes) rotating with the star considering a frame that rotates with the star. Right: The spots with their location (negative longitudes) rotating with the star considering a frame that rotates with the star.}
\label{rotcoordk63}
\end{figure}

\begin{figure}[h]
\centering
\includegraphics[width=8cm]{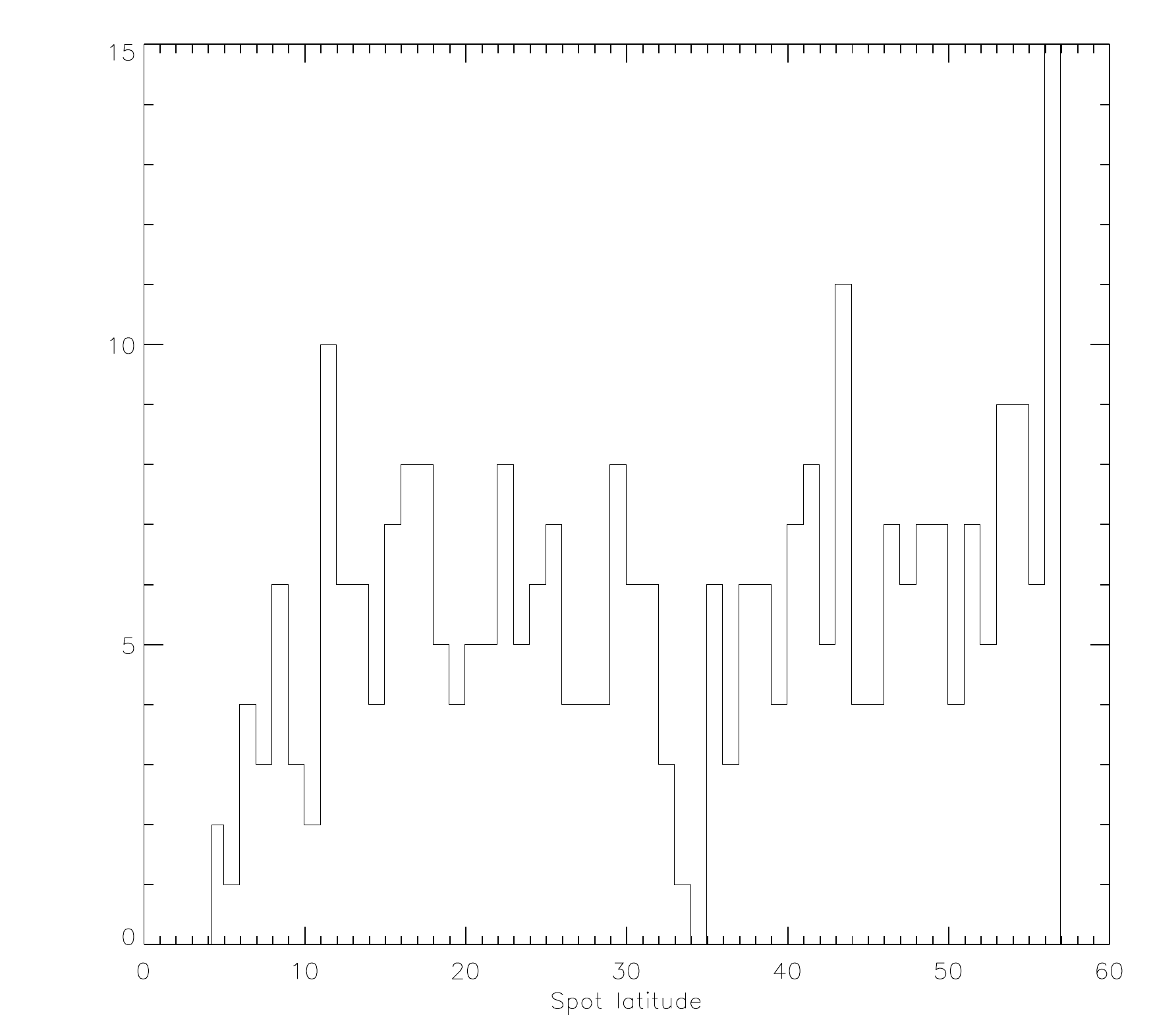} 
     \caption{Latitude distribution of the spots on Kepler-63 showing a gap between low and high distribution.}
\label{fig:histolat}
\end{figure}

Since we assume that the planet transits are in the Northern hemisphere of the star, the highly inclined orbit of the planet forces it to cross stellar latitudes of 4$^\circ$ up to 60$^\circ$. Figure~\ref{rotcoordk63} shows the adopted configuration in this study, where we used a sky-projected stellar obliquity of $\psi = -110^\circ$ and orbital inclination angle of $\theta = 138^\circ$. The left panel shows all 297 the spots for the 4 yrs observation overplotted in a reference frame seen from Earth, while in the middle and right panels the spots coordinates are given in a frame that rotates with the star (positive and negative longitudes), after the application of the rotation matrix. The impact of the values of sky-projected stellar obliquity and inclination angles on our results, in Sect. 4, we shall consider the large uncertainties in the rotation matrix. The latitude distribution of all spots are plotted on Figure~\ref{fig:histolat}, there clearly is a gap at {$\sim 34-35^\circ$} in the latitude distribution, separating the high latitude and the equatorial spots. This might indicate a possible latitudinal bimodality in the spot distribution. \cite{Lehtinen19} found for the young rapidly-rotation solar-type LQ Hya high and low latitude spots with no detection of mid-latitude temperature anomalies. This higher spot activity at the star poles agrees with results obtained by the Doppler imaging technique for young stars. \cite{Strassmeier02} reported that 41 out of 65 stars showed prominent polar spots. Moreover, there is a tendency that polar spots appear preferentially on short-period stars, that is, young stars \citep{Strassmeier04}.

Once each spot longitude and latitude on the stellar rotating reference frame are determined, it is possible to build a diagram of the spots latitude over time. In the case of sunspots, this diagram is known as the ``butterfly diagram", and depicts the high latitude of sunspots at the beginning of a magnetic cycle, and how spots appear closer and closer to the equator as the cycle evolves. This kind of diagram, together with information on differential rotation, is crucial for dynamo models acting on young solar like stars.

\section{Starspots properties and Butterfly diagram}

\begin{figure}[h]
\centering
\includegraphics[width=8cm]{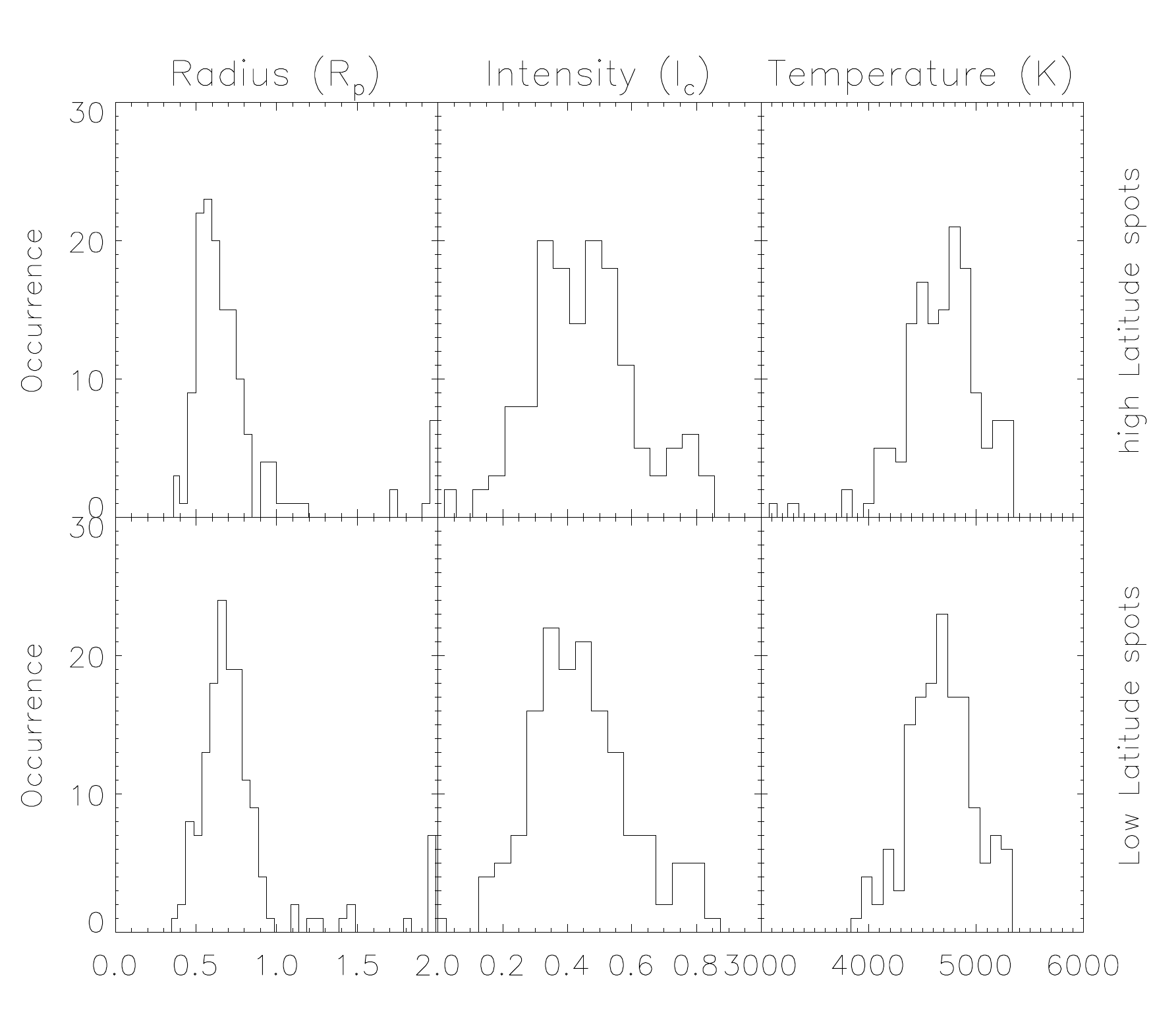} 
     \caption{Histograms of the physical properties of the starspots on Kepler-63. Top panels: High latitude spots. Bottom panels: Low latitude spots. Left: Spot radius in units of planet radius, $R_p$. Middle: Spot intensity in units of disk center intensity, $I_c$. Right: Spot temperature obtained from its intensity.}
\label{histo}
\end{figure}

For all spots, in the high latitude (146 spots) and equatorial (151 spots) distributions, the radius and intensity were inferred from the model fit. Histograms of the physical spot parameters are shown in Figure~\ref{histo}. Assuming that both the stellar surface and the spot emit as black bodies, it is possible to convert from spot intensity, as a fraction of maximum brightness of the stellar disk, $I_c$, to temperature as was done in \cite{SilvaValio10}. The temperature of the spots are shown in the right panels of Figure~\ref{histo}. The average values obtained for the spots radius, intensity, and temperature are given in Table~\ref{spots}.

\begin{table}[h]
\caption{Mean values of the spot parameters on Kepler-63.}
\begin{center}
\begin{tabular}{lc}
\hline\hline
\textit{High latitude spots}\\
\hline
Parameter & Average\\
\hline
Radius, ($R_p$) & $0.76 \pm 0.35$\\
Radius, (Mm) & $31 \pm 14$ \\
Intensity, ($I_c$) & $ 0.48 \pm 0.16$\\
Temperature, (K) & $ 4700 \pm 400 $\\
\\
\textit{Low latitude spots}\\
\hline
Parameter & Average\\
\hline
Radius, ($R_p$) & $0.79 \pm 0.33$\\
Radius, (Mm) & $32 \pm 13$ \\
Intensity, ($I_c$) & $ 0.47 \pm 0.15$\\
Temperature, (K) & $ 4700 \pm 400 $\\
\hline
\end{tabular}
\end{center}
\label{spots}
\end{table}

\begin{figure}[h]
\centering
\includegraphics[width=8cm]{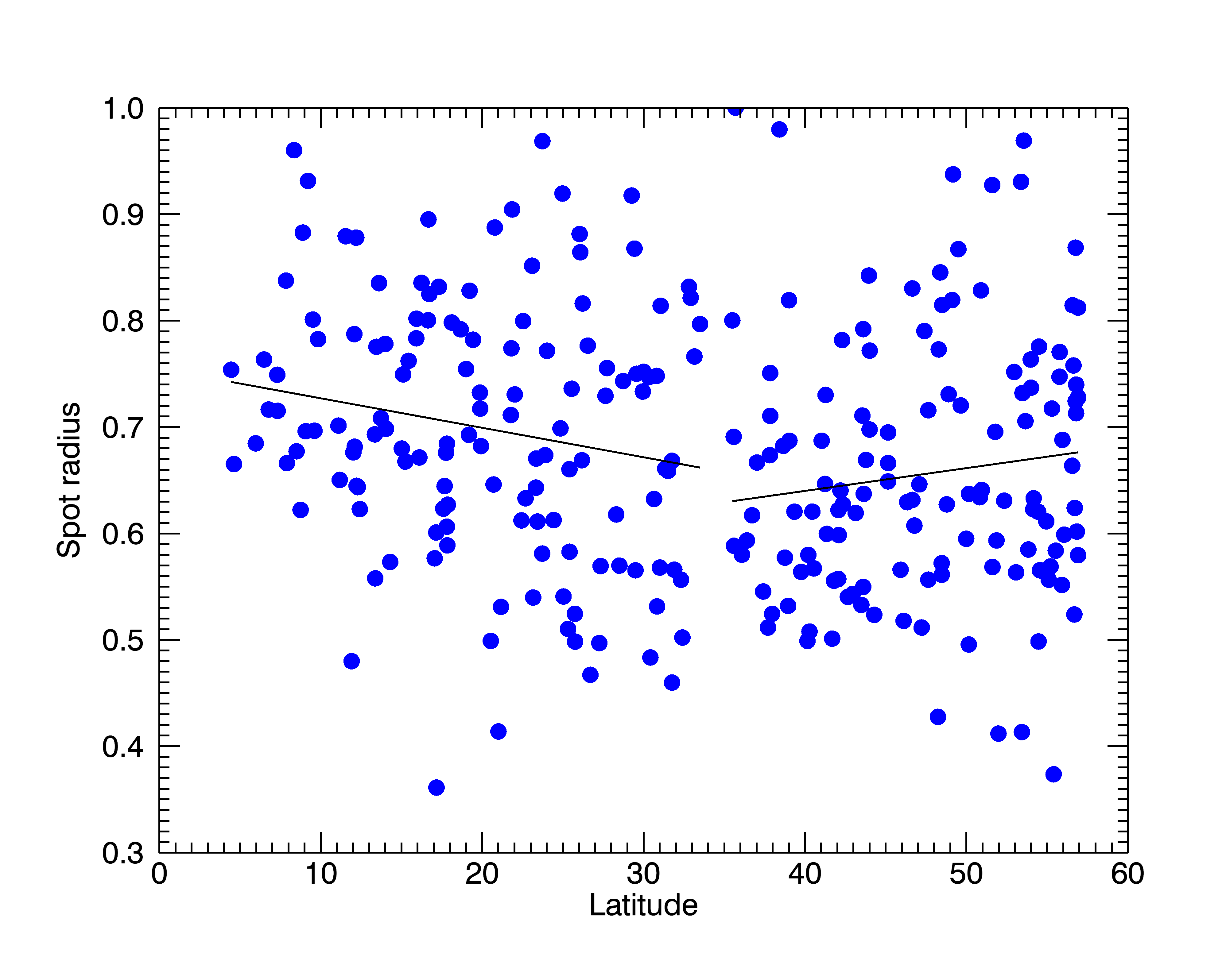} 
     \caption{Spot radius as a function of its latitude, for spots smaller than a planet radius. The black lines represent linear fits to the low ($<34^\circ$) and high ($<34^\circ$) latitude spots.}
\label{fig:radlat}
\end{figure}

We further investigate the existence of different properties of low and high latitude spots by analysing the spots radius with latitude. Figure~\ref{fig:radlat} shows the spots radius for spots with radius smaller than the planetary radius. Separate linear fits (black lines in Figure~\ref{fig:radlat}) were performed to the two spot distributions separated at latitude $34^\circ$. There appears to be a tendency for low latitude spots to be larger close to the equator, whereas the high latitude spots increase in size toward the pole. No such tendency was observed in spot intensity.

\begin{figure}[h]
\centering
\includegraphics[width=8cm]{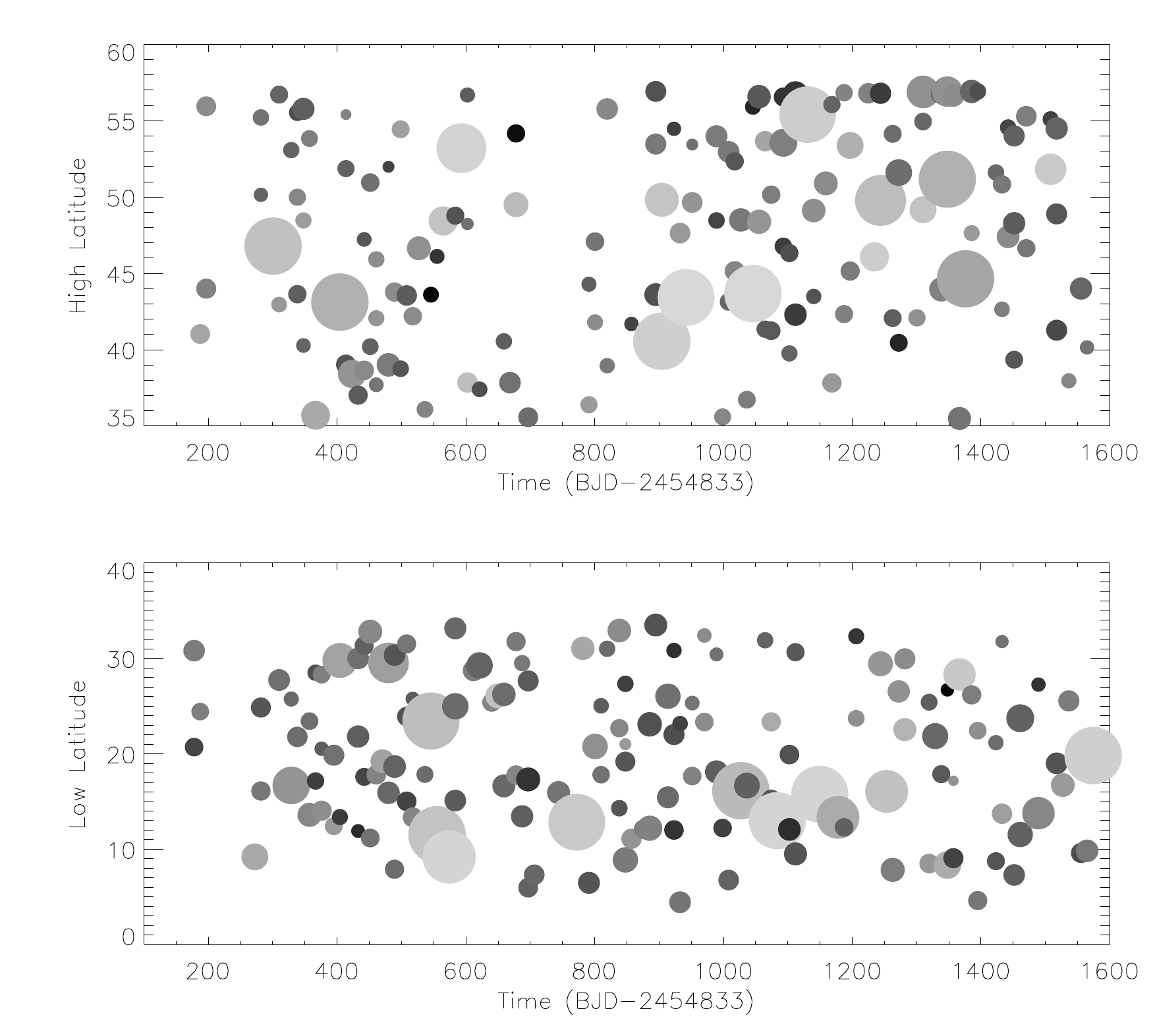} 
     \caption{Kepler-63 "Butterfly diagram": Spots latitude temporal evolution during the 4 years of observation of the Kepler satellite. The size of the disks is proportional to the spot radius, whereas the colour refers to the spots intensity. Top panel: High latitude spots. Bottom panel: Low latitude spots.}
\label{spotdist}
\end{figure}

The latitude of spots as a function of time is presented in  Figure~\ref{spotdist}. It clearly differs from the sunspots distribution, exhibiting spots at latitudes larger than $40^\circ$. The size of the circles is proportional to the spot radius, and its colour refers to the spot intensity.

\begin{figure}[h]
\centering
\includegraphics[width=8cm]{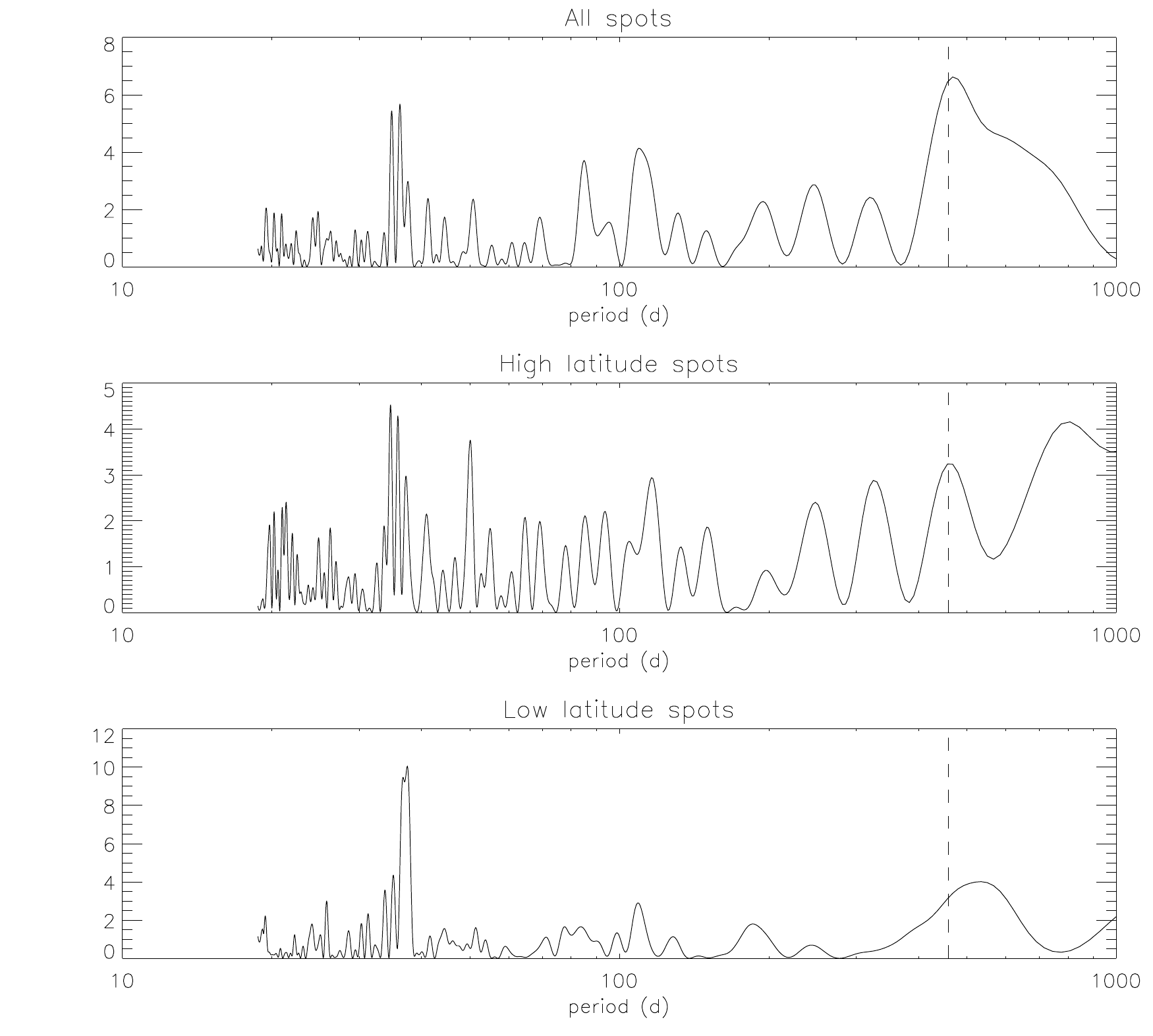} 
     \caption{Lomb-Scargle periodgram of the deficit flux of spots. Top panel: All the 297 spots distribution. Middle panel: Applied for the high latitude distribution. Bottom panel: Applied for the low latitude distribution. The vertical dashed line indicates the value of the magnetic cycle period of 1.27 yr found by \cite{EstrelaValio16}.}
\label{lombscarglespots}
\end{figure}

There exists a degeneracy between the radius and intensity of a spot obtained from the model fit because either a darker small spot or a large brighter spot may produce similar signals in the transit light curve. To avoid this uncertainty, we define the flux deficit of each spot as $D = \pi R_{spot}^2(1 - I_{spot}/I_c)$, where $R_{spot}$ is the spot radius and $I_{spot}/I_c$ the ratio of the spot intensity to the disk centre intensity. Considering this bimodal distribution, we calculated the spots flux deficit for each transit for both distributions. We then applied the Lomb-Scargle periodogram \citep{Lomb76,Scargle82}. In Figure~\ref{lombscarglespots} we have the periodograms for all the spots and for both distributions, where the vertical dashed line is the value for a magnetic cycle period of 1.27 yr or 460 days found for Kepler-63 by \cite{EstrelaValio16}. This magnetic cycle period matches with a peak in the periodogram for the high latitude spots distribution, indicating a possible dominance of the cycle for this distribution. This also may indicate that the spots originate from different processes.

\begin{figure}[h]
\centering
\includegraphics[width=8cm]{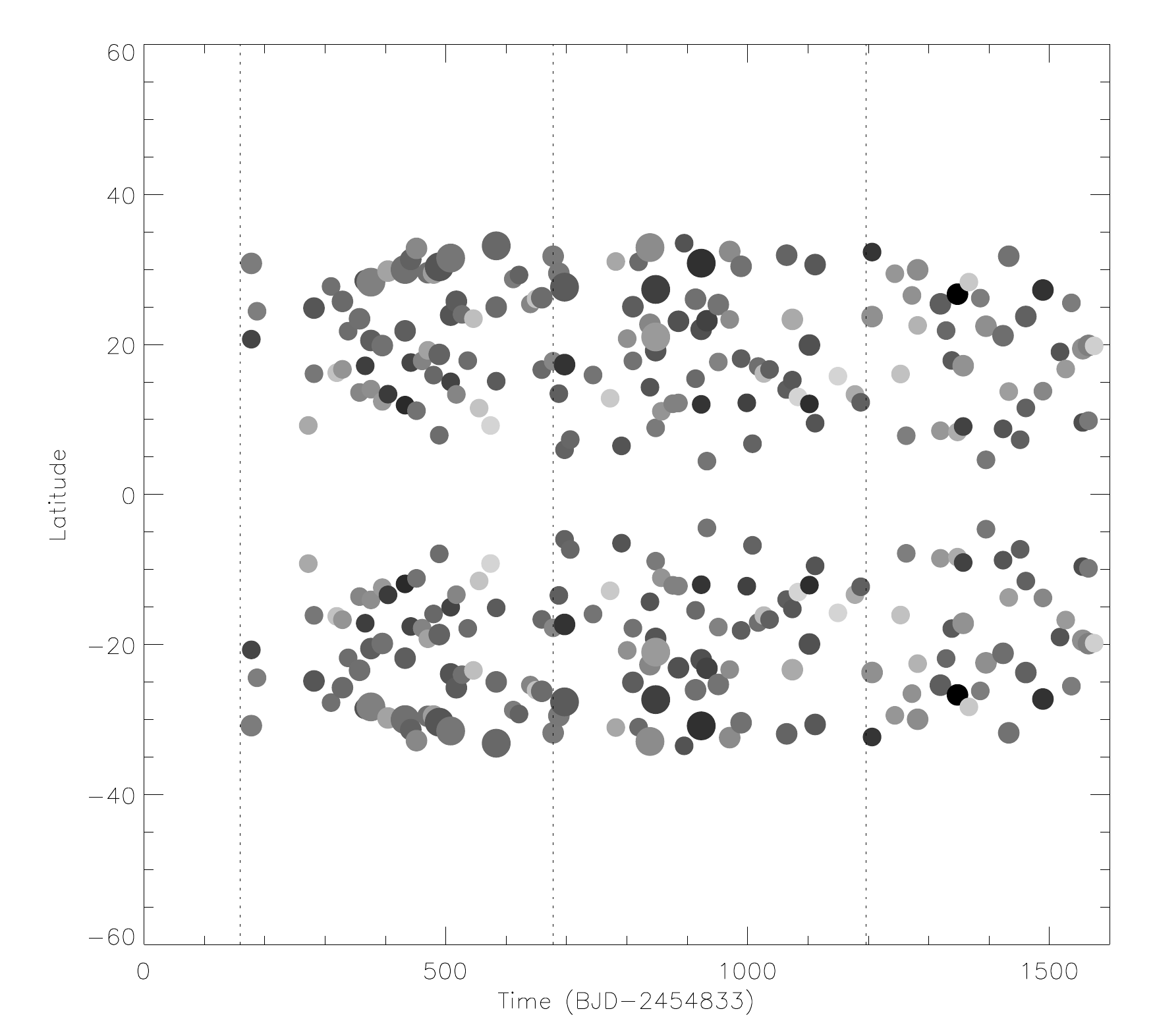} 
     \caption{Butterfly diagram of only the low latitude spots reflected also for negative latitudes to mimic the solar diagram. The vertical dotted are 519d apart, indicating the approximate periods of minimum activity of the magnetic cycle.}
\label{fig:butterflylow}
\end{figure}

To better mimic the solar butterfly diagram, on Figure~\ref{fig:butterflylow} only the low latitude spots are plotted. According to \citet{EstrelaValio16} a short magnetic cycle slightly over a year is at work on Kepler-63. However, the Lomb-Scargle periodogram of the low latitude spots indicated a long term periodicity at around 519d. This is the separation in time of the vertical dotted lines in the figure, guiding the eye to the possible three magnetic activity cycles in action on low latitudes on Kepler-63.

Considering the large uncertainties of the sky-projected obliquity (-124 and -88) and inclination of rotation axis (131 and 145) we calculated the latitudes of starspots considering all the possible scenarios. We still have a bimodal distribution of high latitude and equatorial spots, but the latitudinal gap varies from the latitude $30^\circ$ up to the latitude $46^\circ$. We discarded the values found when using sky-projected obliquity of -88 because it leads to no detection of high latitude spots, which does not agree with the results found by \cite{Ojeda13}.

\section{Differential rotation}\label{difrot}

\begin{figure}[h]
\centering
\includegraphics[width=8cm]{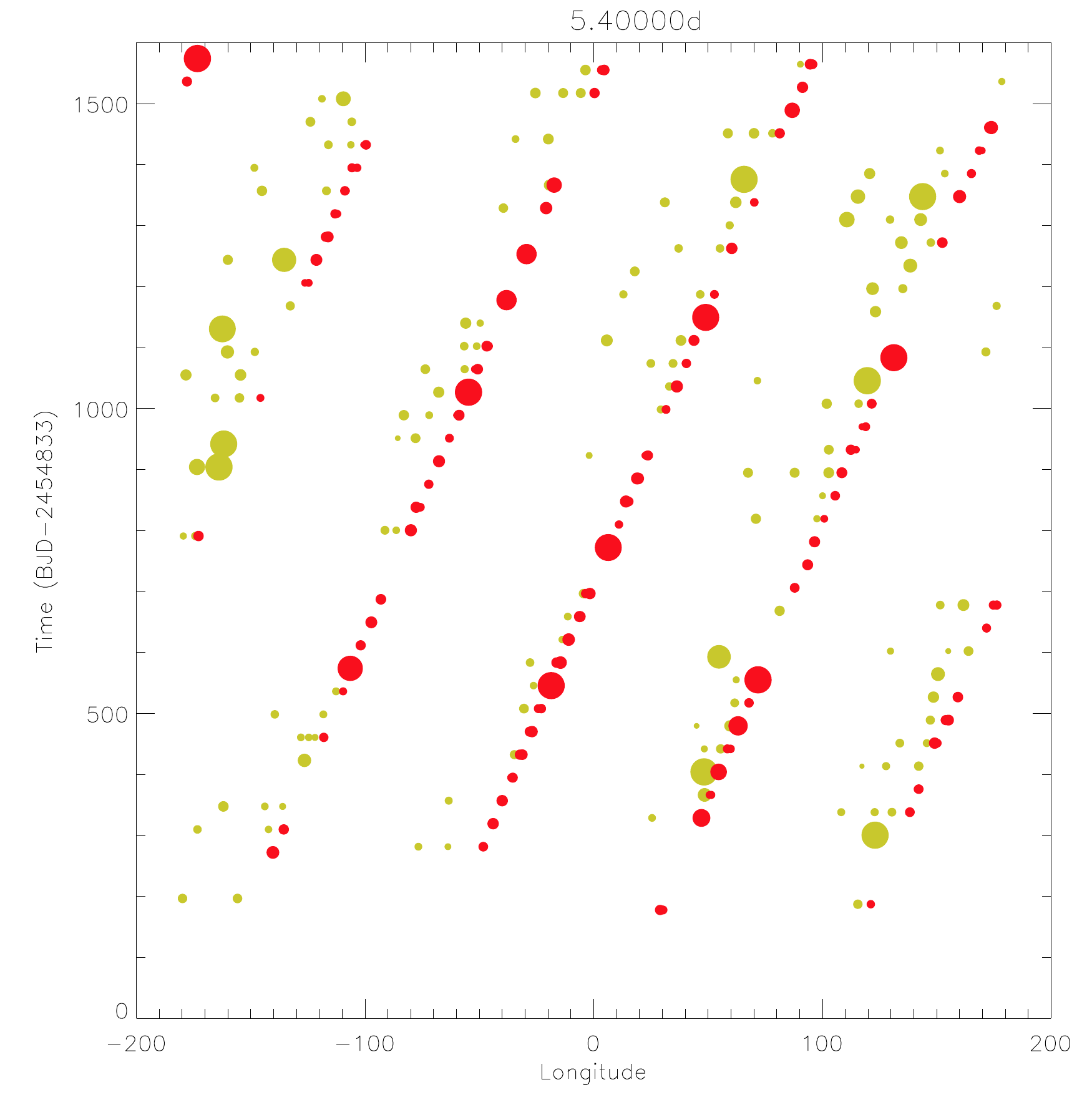} 
     \caption{Longitude of spots in time for all transits measured in a reference frame that rotates with the star with constant $P_{rot} = 5.400$d. Red circles are low latitude spots with $\alpha < 34^\circ$, whereas yellow circles are for high latitude spots ($\alpha < 34^\circ$), where $\alpha$ is the spot latitude.}
\label{fig:spotslg}
\end{figure}

Since the position of the spots is known in every transit, in principle it would be possible to easily determine the differential rotation, as done for other stars using this same model (e.g. CoRoT-2, Kepler-17, and Kepler-71), by analysing the longitude of the spots for each transit.  Figure~\ref{fig:spotslg} shows the longitude of the spots as a function of time for each transit in a reference frame that rotates with the star.. The  circle size in the figure correspond to the  size  of the 297 spots found, considering a period of 5.400d for a rigid rotator. The red circles are for the low latitude spots, whereas the yellow circle are the high latitude spots.

\begin{figure}[h]
\centering
\includegraphics[width=8cm]{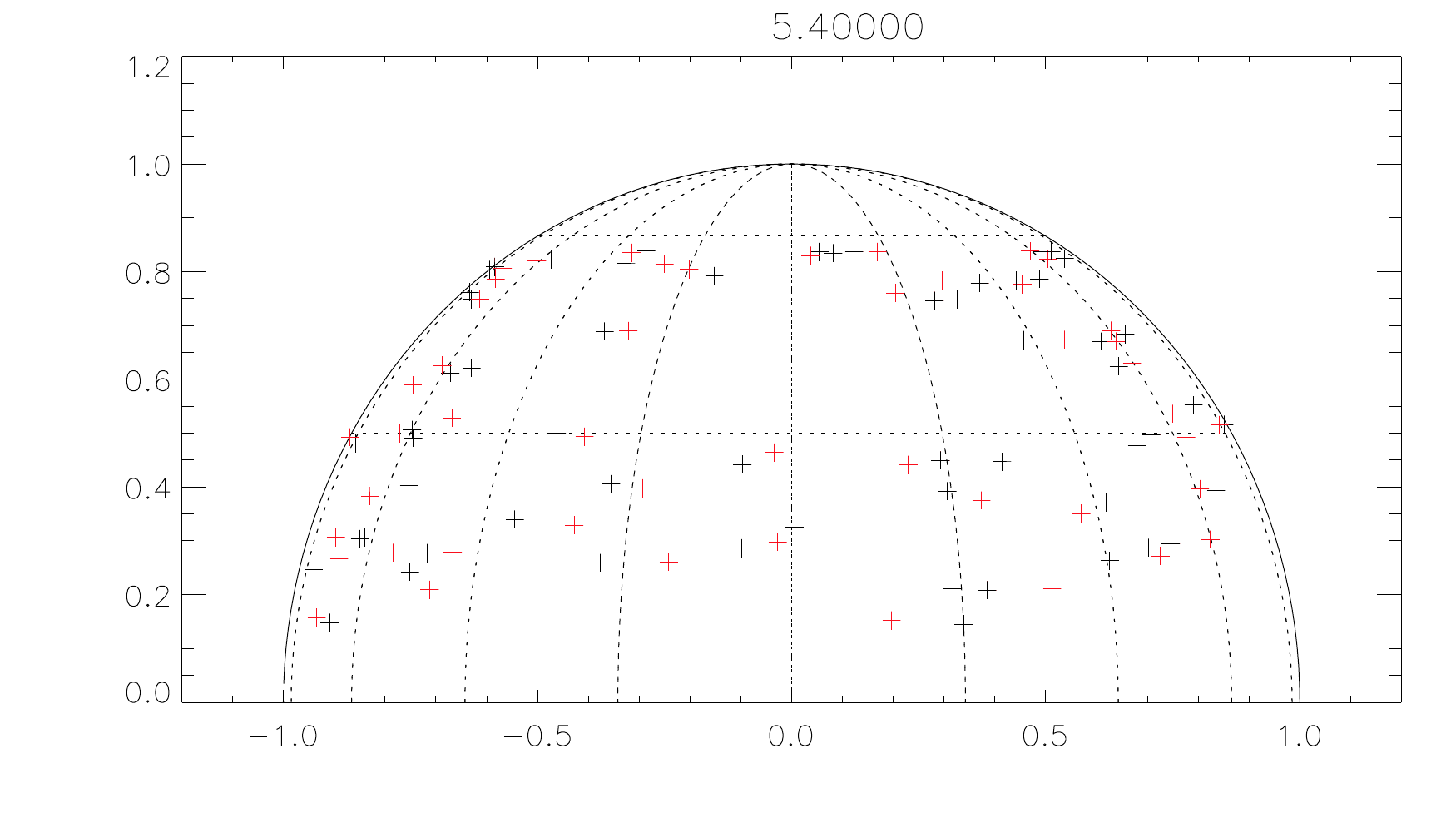}
     \caption{Same spot detection on a later transit considering a stellar rotation period of $P_{rot} = 5.400$d.  Northern hemisphere of Kepler-63 with the 59 spots pairs positions with latitudes within $2^\circ$ and longitudes closer than $10^\circ$ on a later transit (red crosses).}
\label{fig:samespots5400}
\end{figure}

As can be seen in the figure, because of the polar like orbit, only a narrow band in longitude of $\sim13^\circ$ is occulted on each transit, therefore making it very hard to infer differential rotation. Nevertheless, we searched for spots that were detected on a later transit with latitude differences less than $2^\circ$ and longitude within $10^\circ$. A total of 87 pairs of spot satisfied both criteria, however, 28 of these spots were detected more than 75 orbital periods later. Because it is difficult to assess that these are really long lived spots, we consider only spots that were detected less than 10 transits later. Due to the 7:4 resonance between the planet orbital period and the mean rotation period of the star, 31 spots were detected 4 transits later and 28 spots, 8 transits later (lifetimes of $\sim 75.5$d). These 59 pairs of spots are plotted on Figure~\ref{fig:samespots5400} as pairs of  black and red crosses. Next, we estimated the difference in longitude for each pair of spots.

\begin{figure}[h]
\centering
\includegraphics[width=8cm]{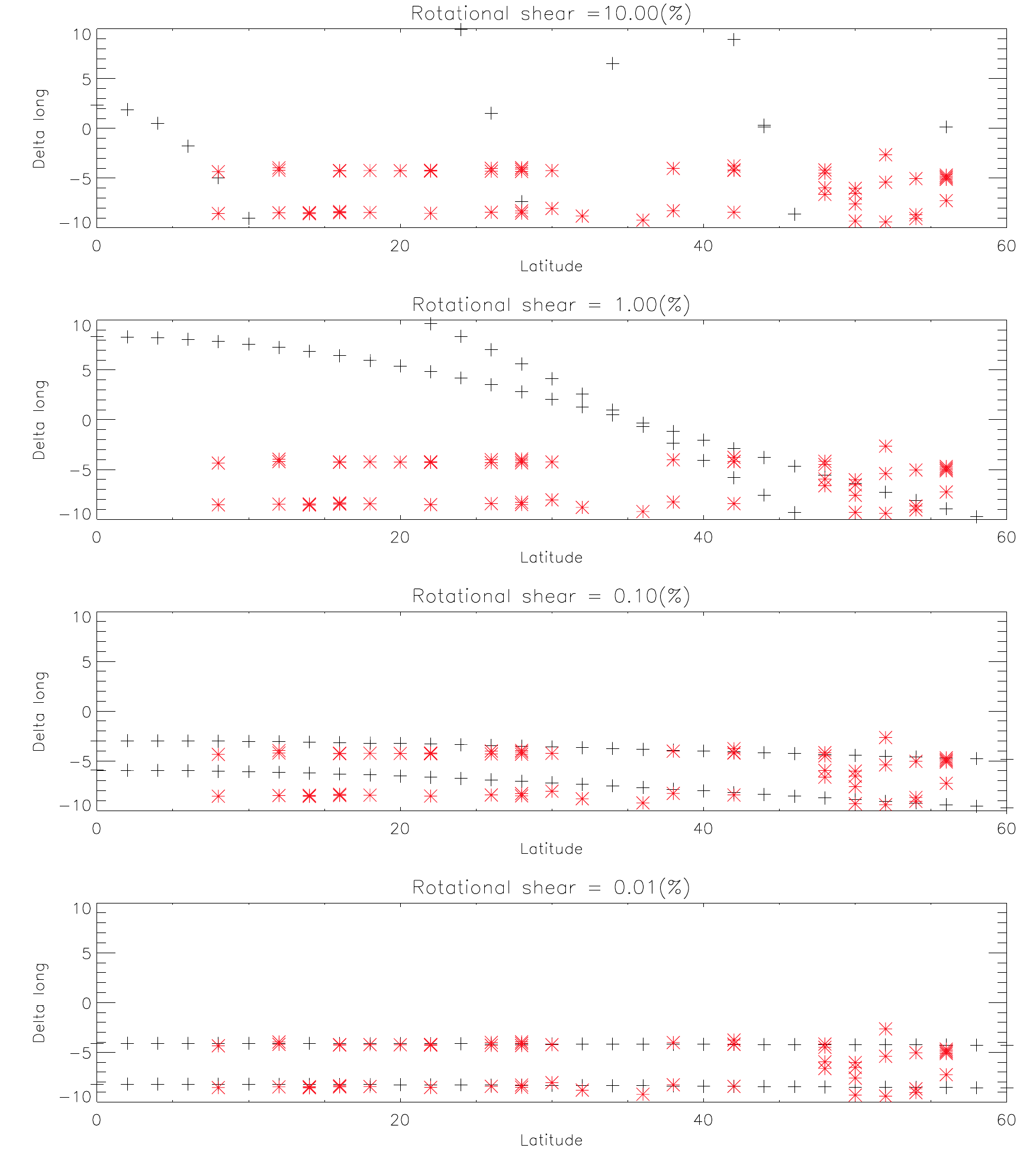}
     \caption{Expected longitude shifts (black crosses) of the same spot detected on a later transit within $10^\circ$, for 4 rotational shear values, from 0.01\% (bottom panel) to 10\% (top panel). The results of the 59 spots detected from transit crossing are plotted as red asterisks.}
\label{fig:dlg}
\end{figure}

Since the longitude of the spots was estimated considering a fixed rotation period, we investigate if it would be possible to measure a high rotational shear. Similar to \citet{SilvaValio08}, the differential rotation is estimated by considering the shift in longitude of the same spot on a later transit. Here we assume a solar-like differential rotation profile: $\Omega(\alpha)$:
\begin{equation}
    \Omega(\alpha) = \Omega_{eq} - \Delta\Omega \sin^2(\alpha)
    \label{eq:difrot}
\end{equation}

\noindent where $\alpha$ is the latitude, $\Omega_{eq}$ is the angular velocity at the stellar equator, and $\Delta\Omega$ is the rotational shear. We simulated the longitude shift in spots after $N$ orbital periods considering a relative rotational shear, $\Delta\Omega / \bar\Omega$, of 0.0001, 0.001, 0.01, and 0.1. Only longitude shifts smaller than $10^\circ$ were considered. The simulated values are plotted as black crosses in Figure~\ref{fig:dlg}, whereas the measured longitude shifts of the 59 detected spots are represented by the red asterisks. As can be seen from the figure, it would be possible to estimate high differential rotation even with a limited longitude band probed during each transit, but only a very small relative rotation shear, $\sim 0.01\%$ (bottom panel), is able to reproduce the observations.

\begin{figure}[h]
\centering
\includegraphics[width=8cm]{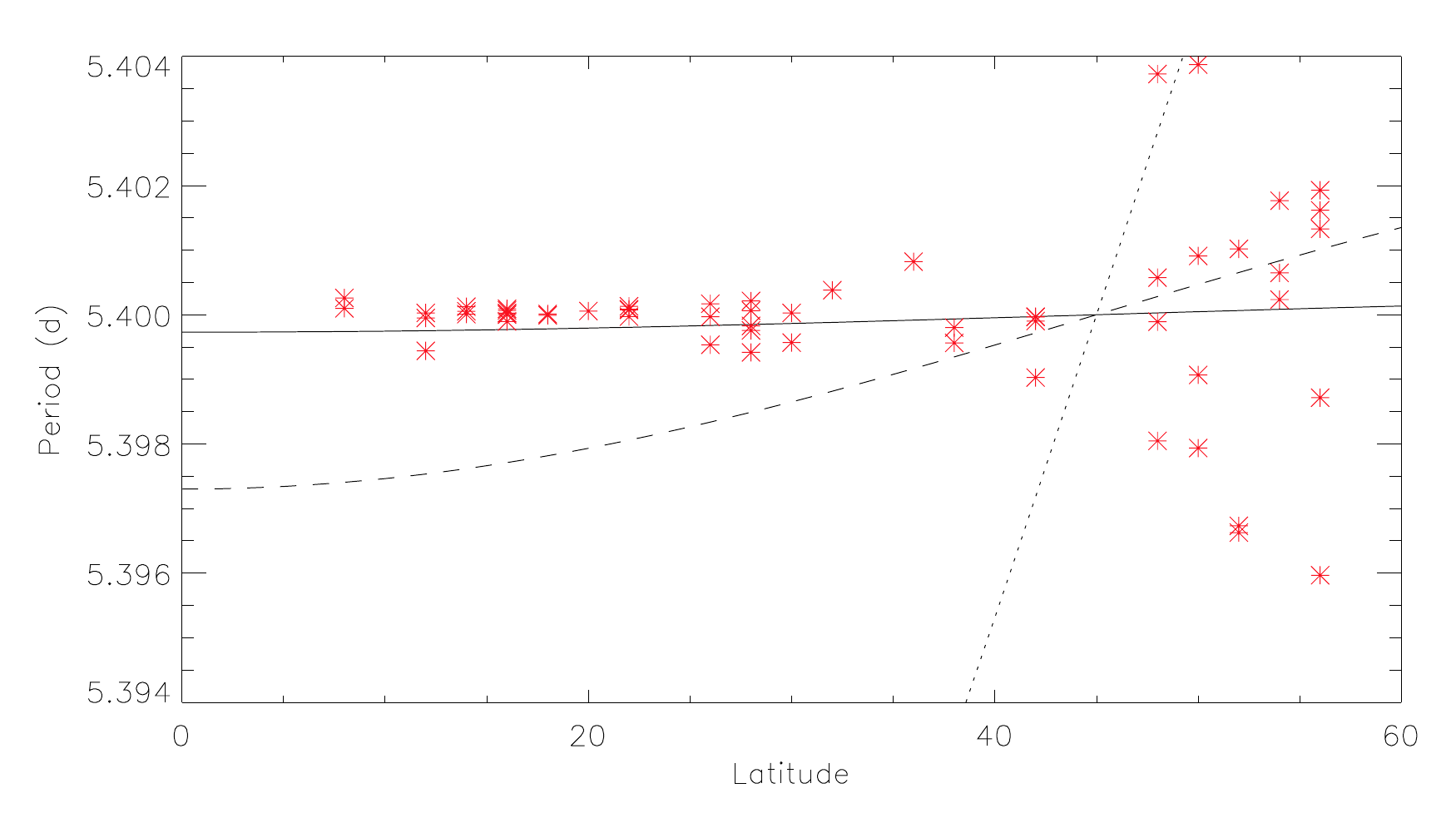} 
     \caption{Stellar rotation period given by Eq.~\ref{eq:pstar} as function of latitude. The black curves represent the rotation profile given by Eq.~\ref{eq:difrot} for $\Delta\Omega /\bar\Omega$ = 0.0001 (solid), 0.001 (dashed), and 0.01 (dotted).}
\label{fig:difrot}
\end{figure}

The angular velocity, $\Omega(\alpha)$, for each latitude is given by $\Omega = 2\pi / P_{star}(\alpha)$, where $P_{star}(\alpha)$ is calculated according to 
\begin{equation}
    P_{star}(\alpha) = 360^\circ N P_{orb}  \left(360^\circ {N P_{orb} \over P_{rot}} - \Delta long(\alpha)\right)^{-1}
    \label{eq:pstar}
\end{equation}
  
\noindent where $P_{rot}=5.400$d is the mean stellar rotation period, $P_{orb}$, the planet orbital period, $\Delta long(\alpha)$ is the difference in longitudes calculated for the same spot identified at approximately the same position $N$ transits later. Using the longitude differences, $\Delta long(\alpha)$, plotted on Figure~\ref{fig:dlg}, we obtain the rotation profile calculated using Eq.~\ref{eq:pstar} of Kepler-63, that is plotted on Figure~\ref{fig:difrot} as red asterisks. Also plotted are the rotation profile for a relative rotation shear of 0.01\% (solid curve), 0.1\% (dashed curve), and 1\% (dotted curve).

Again the smallest rotation shear of 0.01\% is the one that best represents the data, at least for latitudes smaller than $40^\circ$, yielding  $\Omega_{eq} = 1.16$ rd/d and a rotation shear of $\Delta\Omega = 0.000116$ rd/d, or a relative differential rotation of 
$\Delta\Omega/\bar\Omega = 0.01$ \%, where $\bar\Omega = 2\pi / P_{rot}$ is the mean angular velocity. 

%The rotation profile of Kepler-63, that is the rotation period with latitude, is plotted in Figure~\ref{fig:difrot}, where the asterisks represent the data and the red solid curve is the solar-like fit for latitudes less than $34^\circ$. Due to the discrepant behaviour of the rotation periods for low and high latitude spots, only the rotation of low latitude spots were fit.

%_________________________________________________________________
\section{Discussion and Conclusions}

The aim of this work was to characterise the spots of a young solar type star, its parameters and distribution on the stellar surface, and also its rotation pattern. To build the butterfly diagram of the star, spots at different latitudes are needed. Thus we searched the database of the transiting planets of Kepler stars for those which transit light curve showed variations of a few percent indicating spot activity on their surfaces. Kepler-63, due to the highly inclined orbit of its planet, seemed the most suited star to be analysed. Kepler-63 was observed for 1415 days, during which a total of 150 transits were detected. Its light curve showed modulations of up to 4\% peak-to-peak variation. Here we extended the analysis done by \cite{Ojeda13} for the whole time of observation of Kepler-63 (about 4 years), exploring the distribution of the spots. Especially, we calculated the spot position, longitude and latitude, on a reference frame that rotates with the star. Moreover, we also estimated the differential rotation of the star.

Here we analysed the star Kepler-63, that hosts a planet in an almost polar orbit. This fortuitous geometry allowed us to map the spots on the surface of the star at different stellar latitudes. The presence of spots on the stellar surface influence the transit light curve in two ways which can cause a different determination of the orbital parameters of a planet in orbit \citep{SilvaValio10}. We found slightly increased values from \cite{Ojeda13} for the planet radius of 0.0644 $R_{star}$ and semi major axis $a = 19.35 R_{star}$. We then applied the method developed by \cite{Silva03}, resulting in a total of 297 spots that were detected in the 150 observed transits, and their physical characteristics inferred. 
The location of the spots in a reference frame that rotates with the star was determined by the application of a rotation matrix that considers the sky-projected stellar obliquity of $-110^\circ$, inclination of rotation axis of $138^\circ$, and rotation $\Omega t$.

The distribution of spots latitude indicated a possible bimodality with a gap at around $34^\circ$. About half of the spots are located near the poles of the star, as expected for young stars. \cite{Ojeda13} had already indicated the presence of polar spots on Kepler-63. Our analyses suggest that high latitude spots distribution dominate the magnetic cycle period of Kepler-63, also indicating that the spots from high and low latitude originate from different processes (see Figures~\ref{fig:radlat}, \ref{lombscarglespots} and \ref{fig:spotslg}). Moreover, low latitude spots seem to increase in size toward the equator, whereas the high latitude ones have larger radius when closer to the poles.
 
CoRoT-2 and Kepler-63 are young stars of similar age with less than 500 Myr, while Kepler-71 and the Sun are much older, with ages of 2.5-4.0 and 4.6 Gyr, respectively. The spectral type of Kepler-63 is not well determined, nevertheless its effective temperature, (5576 $\pm$ 50)K \citep{Ojeda13}, is similar to CoRoT-2 (5575 $\pm$ 66)K \citep{Torres12}, Kepler-17 (5781K) \citep{Valio17}, Kepler 71 (5540 $\pm$ 120)K \citep{Zaleski19}, and the Sun (5778 $\pm$ 3)K \citep{Stix02}. 

\begin{table}[h]
\caption{Mean values of the spot parameters on Kepler-63, other stars, and the Sun.}
\fontsize{5.8pt}{9.25pt}\selectfont
\begin{tabular}{lccccc}
\hline\hline
Star & Kepler-63$^{(1)}$ & Kepler-17$^{(2)}$ & Kepler-71$^{(3)}$ & CoRoT-2$^{(4)}$ & Sun$^{(5)}$            \\
\hline
Mass ($M_{Sun}$)    & 0.984   & 1.16       & 0.997          & 0.97           & 1.0            \\
Radius ($R_{Sun}$)  & 0.901  & 1.05        & 0.887          & 0.902          & 1.0            \\
$T_{eff}$ (K)       & 5576   & 5781        & 5540           & 5575           & 5778           \\
Age (Gyr)           & 0.2  & $>1.8$     & 2.5-4.0        & 0.13-0.5       & 4.6            \\
$P_{rot}$ (d) & 5.405 & 12.28 & 19.77 & 4.54 & 27.6 \\
$\Omega$ (rd/d) & 0.000116 & 0.077 & 0.005 & 0.042 & 0.50 \\
$\Delta\Omega/\Omega$ (\%) & 0.01 & 15.0 & 1.8 & 3.0 & 22.1 \\
\\
Planet & Kepler-63b & Kepler-17b & Kepler-71b & CoRoT-2b &    \\
\hline
Radius ($R_p/R_{star}$) & 0.0644 &   0.138       & 0.1358         & 0.172          &                \\
a ($a/R_{star}$)      & 19.35  &  2.45       & 12.186         & 6.7            &                \\
\\
Starspots\\                                                         \\
\hline
Radius (Mm)         & $32 \pm 14$ & $80 \pm 50$    & $51 \pm 26$    & $55 \pm 19$    & $12 \pm 10$    \\
$T_{spot}$ (K)      & $4700 \pm 300$ & $5100 \pm 500$ & $4800 \pm 500$ & $4600 \pm 700$ & $4800 \pm 400$ \\
\hline
\end{tabular}
\label{starspots}
\tablebib{(1):~\citet{Ojeda13} and present study; (2):~\citet{Valio17}; (3):~\citet{Zaleski19}; (4):~\citet{SilvaValio10, SilvaValioLanza11}; (5):~\citet{Stix02}.}
\end{table}

The physical parameters of Kepler-63, other solar-like stars, and the Sun are shown in Table~\ref{starspots}. Kepler-63 and CoRoT-2 are much more active and younger than the Sun but present similar average spot temperature, $T_{spot}$. 
CoRoT-2, Kepler-17, and Kepler-71 were analysed using the same method applied in this work. Since these star are really similar to Kepler-63, particularly CoRoT-2 of similar age, we might expect them to have spots with same size but Kepler-63 presents smaller spots. Because this method uses a transiting planet as a probe of the contrasting features in the stellar surface, it is more precise for a smaller planet. CoRoT-2b is almost 3 times the size of Kepler-63b, which agrees with the limitations we have when characterising starspots. 

Kepler-63 was found to rotate almost rigidly for latitudes less than $34^\circ$, but presenting no organised pattern for larger latitudes. From the stars analysed so far with this method this is the star with the smaller differential rotation. Moreover, long lived spots with lifetimes of at least 75 days were also found. Such long spot lifetimes are expected on stars with low surface shear.

Applying the method of starspot transit mapping, developed by \cite{Silva03}, it was possible to produce a latitudinal distribution of spots and a ``butterfly diagram". The indication of a possible bimodality in the spots latitude distribution with different processes at work for the origin of the low and high latitude spots as well as spot lifetime and the differential rotation profile are important key ingredients related to dynamo models. Applying this method we hope to have contributed to a better understanding of the stellar dynamo. 

%_________________________________________________________________
\begin{acknowledgements}

We thank the anonymous referee for comments and suggestions which improved the original work.
We are grateful to the Brazilian funding agency FAPESP (\#2013/10559-5). We also gratefully acknowledge the support from the Brazilian Federal Agency for Support and Evaluation of Graduate Education (CAPES).

\end{acknowledgements}
%-----------------------------------------------------------------


\begin{thebibliography}{}
\bibitem[Barnes et al.(2005)]{Barnes05}  Barnes, J. R., Collier Cameron, A., Donati, J.-F., James, D. J., Marsden, S. C. \& Petit, P. 2005, \mnras, 357, L1

\bibitem[B{\'e}ky et al.(2014)]{Beky14} B{\'e}ky, B., Kipping, D.~M., \& Holman, M.~J.\ 2014, \mnras, 442, 3686

\bibitem[Berdyugina(2005)]{Berdyugina05} Berdyugina, S.~V.\ 2005, Living Reviews in Solar Physics, 2,  

\bibitem[Brown et al.(2001)]{Brown01} Brown, T.~M., Charbonneau, D., Gilliland, R.~L., et al.\ 2001, \apj, 552, 699

\bibitem[Christiansen et al.(2012)]{Christiansen12} Christiansen, J.~L., Jenkins, J.~M., Caldwell, D.~A., et al.\ 2012, \pasp, 124, 1279

\bibitem[Estrela \& Valio(2016)]{EstrelaValio16} Estrela, R. \& Valio, A.\ 2016, \apj, 831, 57.

\bibitem[Hackman et al.(2019)]{Hackman19} Hackman, T., Ilyin, I., Lehtinen, J.~J., et al.\ 2019, \aap, 625, A79

\bibitem[Hall, \& Busby(1990)]{Hall90} Hall, D.~S., \& Busby, M.~R.\ 1990, NATO Advanced Science Institutes (ASI) Series C, 377

\bibitem[Hall, \& Henry(1994)]{Hall94} Hall, D.~S., \& Henry, G.~W.\ 1994, International Amateur-Professional Photoelectric Photometry Communications, 55, 51

\bibitem[Herrero et al.(2016)]{Herrero16} Herrero, E., Ribas, I., Jordi, C., et al.\ 2016, \aap, 586, A131

\bibitem[Juvan et al.(2018)]{Juvan18} Juvan, I.~G., Lendl, M., Cubillos, P.~E., et al.\ 2018, \aap, 610, A15

\bibitem[Lanza et al.(2019)]{Lanza19} Lanza, A.~F., Netto, Y., Bonomo, A.~S., et al.\ 2019, \aap, 626, A38

\bibitem[Lehtinen et al.(2019)]{Lehtinen19} Lehtinen, J.~J., K{\"a}pyl{\"a}, M.~J., Hackman, T., et al.\ 2019, arXiv e-prints, arXiv:1909.11028

\bibitem[Lomb(1976)]{Lomb76} Lomb, N.~R.\ 1976, \apss, 39, 447

\bibitem[Maxted(2016)]{Maxted16} Maxted, P.~F.~L.\ 2016, \aap, 591, A111

\bibitem[Montalto et al.(2014)]{Montalto14} Montalto, M., 
Bou{\'e}, G., Oshagh, M., et al.\ 2014, \mnras, 444, 1721

\bibitem[Morris et al.(2017)]{Morris17} Morris, B.~M., Hebb, L., Davenport, J.~R.~A., et al.\ 2017, \apj, 846, 99

\bibitem[Oshagh et al.(2013)]{Oshagh13} Oshagh, M., Boisse, I., Bou{\'e}, G., et al.\ 2013, \aap, 549, A35

\bibitem[Parker(1955)]{Parker55} Parker, E.~N.\ 1955, \apj, 122, 293

\bibitem[Sanchis-Ojeda, \& Winn(2011)]{SanchisOjeda11} Sanchis-Ojeda, R., \& Winn, J.~N.\ 2011, \apj, 743, 61

\bibitem[Sanchis-Ojeda et al.(2013)]{Ojeda13} Sanchis-Ojeda, R., Winn, J.~N., Marcy, G.~W., et al.\ 2013, \apj, 775, 54

\bibitem[Scargle(1982)]{Scargle82} Scargle, J.~D.\ 1982, \apj, 263, 835 

\bibitem[Schrijver \& Zwaan(2000)]{Schrijver00} Schrijver, C.~J., \& Zwaan, C.\ 2000, Solar and stellar magnetic activity / Carolus J.~Schrijver, Cornelius Zwaan.~ New York : Cambridge University Press, 2000.~(Cambridge astrophysics series ; 34),

\bibitem[Silva(2003)]{Silva03} Silva, A.~V.~R.\ 2003, \apjl, 585, L147

\bibitem[Silva-Valio(2008)]{SilvaValio08} Silva-Valio, A.\ 2008, \apjl, 683, L179

\bibitem[Silva-Valio et al.(2010)]{SilvaValio10} Silva-Valio, A., Lanza, A.~F., Alonso, R., \& Barge, P.\ 2010, \aap, 510, A25 

\bibitem[Silva-Valio \& Lanza(2011)]{SilvaValioLanza11} Silva-Valio, A., \& Lanza, A.~F.\ 2011, \aap, 529, A36

\bibitem[Stix(2002)]{Stix02} Stix, M.\ 2002, The sun : an introduction -- 2nd ed.~/Michael Stix.~Berlin : Springer, 2002.~QB 521 .S75, 

\bibitem[Strassmeier(2002)]{Strassmeier02} Strassmeier, K.~G.\ 2002, Astronomische Nachrichten, 323, 309

\bibitem[Strassmeier(2004)]{Strassmeier04} Strassmeier, K.~G.\ 2004, Stars as Suns : Activity, Evolution and Planets, 219, 11 

\bibitem[Strassmeier(2009)]{Strassmeier09} Strassmeier, K.~G.\ 2009, \aa, 17, 251 

\bibitem[Stumpe et al.(2012)]{Stumpe12} Stumpe, M.~C., Smith, J.~C., Van Cleve, J.~E., et al.\ 2012, \pasp, 124, 985

\bibitem[Stumpe et al.(2014)]{Stumpe14} Stumpe, M.~C., Smith, J.~C., Catanzarite, J.~H., et al.\ 2014, \pasp, 126, 100

\bibitem[Thomas \& Weiss(2008)]{Thomas08} Thomas, J.~H., \& Weiss, N.~O.\ 2008, Sunspots and Starspots,  New York: Cambridge University Press.

\bibitem[Torres et al.(2012)]{Torres12} Torres, G., Fischer, D.~A., Sozzetti, A., et al.\ 2012, \apj, 757, 161 
\bibitem[Tregloan-Reed et al.(2013)]{Tregloan13} Tregloan-Reed, J., Southworth, J., \& Tappert, C.\ 2013, \mnras, 428, 3671

\bibitem[Tregloan-Reed et al.(2015)]{Tregloan15} Tregloan-Reed, J., Southworth, J., Burgdorf, M., et al.\ 2015, \mnras, 450, 1760

\bibitem[Tregloan-Reed et al.(2018)]{Tregloan18} Tregloan-Reed, J., Southworth, J., Mancini, L., et al.\ 2018, \mnras, 474, 5485

\bibitem[Valio(2013)]{Valio13} Valio, A.\ 2013, New Quests in Stellar Astrophysics III: A Panchromatic View of Solar-Like Stars, With and Without Planets, 472, 239

\bibitem[Valio et al.(2017)]{Valio17} Valio, A., Estrela, R., Netto, Y., et al.\ 2017, \apj, 835, 294

\bibitem[Zaleski et al.(2019)]{Zaleski19} Zaleski, S.~M., Valio, A., Marsden, S.~C., et al.\ 2019, \mnras, 484, 618

\end{thebibliography}
\end{document}